\documentclass[aps,prb,superscriptaddress,twocolumn,longbibliography]{revtex4-1}
\usepackage[utf8]{inputenc}
\usepackage[english]{babel}
\usepackage{amsmath}
\usepackage{hyperref}
\usepackage{textcomp} 
\usepackage{amsfonts}
\usepackage{amssymb}
\usepackage{graphicx}
\newcommand{\ital}[1]{\section{#1}}
\newcommand{\mbf}{\mathbf} 
\renewcommand{\k}{{\mbf k}}

\newcommand{\q}{{\mbf q}}

\begin{document}
\title{Orbital selective pairing and gap structures of iron-based superconductors}

\author{Andreas Kreisel}
\affiliation{Niels Bohr Institute, University of Copenhagen, Juliane Maries Vej 30, DK 2100 Copenhagen, Denmark}
\affiliation{Institut f\" ur Theoretische Physik, Universit\" at Leipzig, D-04103 Leipzig, Germany}
\author{Brian M. Andersen}
\affiliation{Niels Bohr Institute, University of Copenhagen, Juliane Maries Vej 30, DK 2100 Copenhagen, Denmark}
\author{P.O. Sprau}
\affiliation{LASSP, Department of Physics, Cornell University, Ithaca, NY 14853, USA}
\affiliation{CMPMS Department, Brookhaven National Laboratory, Upton, NY 11973, USA}
\author{A. Kostin}
\affiliation{LASSP, Department of Physics, Cornell University, Ithaca, NY 14853, USA}
\affiliation{CMPMS Department, Brookhaven National Laboratory, Upton, NY 11973, USA}
\author{J.C. S\'eamus Davis}
\affiliation{LASSP, Department of Physics, Cornell University, Ithaca, NY 14853, USA}
\affiliation{CMPMS Department, Brookhaven National Laboratory, Upton, NY 11973, USA}
\author{ P. J. Hirschfeld}
\affiliation{Department of Physics, University of Florida, Gainesville, FL 32611, USA}

\begin{abstract}
 We discuss  the influence on spin-fluctuation pairing theory of orbital selective strong correlation effects in Fe-based superconductors, particularly Fe chalcogenide systems.
 We propose that a key ingredient for an improved itinerant pairing theory is orbital selectivity, i.e., incorporating the reduced coherence of quasiparticles occupying specific orbital states. This modifies the usual spin-fluctuation via
 suppression of pair scattering processes involving those less coherent states and results in orbital selective Cooper pairing of electrons in the remaining  states.  We show that this paradigm
yields remarkably good agreement with the experimentally observed  anisotropic  gap structures in both bulk and monolayer
FeSe, as well as LiFeAs, indicating that orbital selective Cooper pairing plays a key role in the more strongly correlated iron-based superconductors.
  \end{abstract}
\maketitle

{\ital{Introduction}}\vspace{-0.4cm}
In both copper-based and iron-based high temperature superconductors,  fundamental issues include  the degree of electron correlation and  its  consequences  for enhancing  superconductivity. In both archetypes, there are multiple active orbitals (two O $p$ orbitals and one Cu $d$ orbital in the former, and five Fe $d$ orbitals in the latter). This implies the possibility of orbital-selective physics, where states dominated by electrons of one orbital type may be weakly correlated and others  much more strongly correlated, leading to substantial differences in quasiparticle spectral weights, interactions,  magnetism and orbital ordering\cite{deMedici_review,Bascones_review,Biermann_review,Yin2011,Li_orb_sel_16,Haule16,Ye_doping_FeSc_14}. Cooper pairing itself could then become orbital-selective, \cite{Ogata_selectivepairing,Si_selectivepairing} with the electrons of a specific orbital character binding to form the Cooper pairs of the superconductor. The superconducting energy gaps of such a material would therefore generically be highly anisotropic \cite{Ogata_selectivepairing,Si_selectivepairing}, i.e., large only for those Fermi surface regions where a specific orbital character dominates. Such phenomena, although long the focus of theoretical research on higher temperature superconductivity in correlated multi-orbital superconductors, have remained largely unexplored because orbital-selective Cooper pairing 
has not been experimentally accessible.
 
 \begin{figure*}[tb]
	 \includegraphics[width=\linewidth]{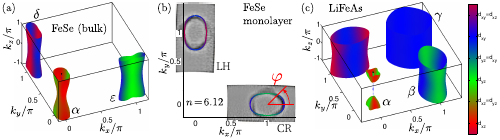}
\caption{Fermi surfaces together with orbital character of the models considered in this work obtained from tight-binding models fit to ARPES and  quantum oscillation experiments. The individual sheets are labeled as indicated: (a) model for FeSe (bulk)\cite{Davis_OSP} including orbital order, (b) 2D model for FeSe monolayer derived from the previous one where maps of ARPES intensities obtained from measurements with horizontally polarized (LH) and circular polarized (CR) initial photons have been overlayed to show agreement to experimental results\cite{Zhang16} and (c) model for LiFeAs\cite{Wang13}. Plots as a function of the angle $\varphi$ around the Fermi surface sheets are done with the angle measured from the $k_x$ axis as indicated in (b).  
\label{fig_fermi}}
\end{figure*}

 Spin fluctuations  are proposed as  the dominant mechanism driving Cooper pairing
 in a wide variety of  unconventional superconductors: heavy-fermion systems, cuprates, two-dimensional organic
charge transfer salts, and iron-based
superconductors (FeSC)\cite{Scalapino2012,Ardavan2012,HirschfeldCRAS,Chubukov_review}.
There is currently no version of spin-fluctuation based pairing theory that enjoys either the well-controlled 
derivation from fundamental interactions or the consensual success explaining observed properties of the BCS-Migdal-Eliashberg theory of 
conventional superconductivity. On the other hand, the calculational scheme referred to as random
phase approximation (RPA) in the case of one-band systems\cite{BerkSchrieffer,ScalapinoLohHirsch86}, or matrix-RPA in the case of multi-band systems\cite{Takimoto2004,Altmeyer2016},
has achieved considerable qualitative progress for unconventional systems.

While material-specific calculations of the critical temperature $T_c$ within spin-fluctuation theory
appear distant,  considerable success has been achieved understanding qualitative aspects of pairing, particularly in 
Fe-pnictide systems\cite{HKM_ROPP,HirschfeldCRAS,Thomale_review}.    In the 122 materials, which were the subject of the
most intensive early study,  itinerant spin-fluctuation theory provided convincing, material-specific  understanding of  the variation
of gap anisotropy with doping within the dominant sign-changing $s$-wave channel, particularly the existence or nonexistence of nodes;  the interplay with $d$-wave pairing;  
the rough size of $T_c$; and the origin of particle-hole asymmetry in the phase diagram. In retrospect, such agreement was somewhat fortuitous,  possibly because
 the 122 systems have large Fermi surface pockets of both hole- and electron-type, and are relatively weakly correlated.   In other pnictides like 111\cite{Yin2011,Ferber2012,Lee_etal_Kotliar2012}, and in 11 Fe-chalcogenide systems\cite{Si_review,Biermann_review}, correlation effects are considerably more 
      significant.  In LiFeAs, for example, angle-resolved photemission spectroscopy (ARPES) measurements\cite{BorisenkoLiFeAs,Borisenko12} show that the $\Gamma$-centered $d_{xz}/d_{yz}$ hole pockets are considerably smaller than predicted by density functional theory (DFT),
      while the $d_{xy}$ pocket is larger. Taking these effects into account via a set of renormalized energy bands is insufficient, however, to account for the
      accurate gap structure of LiFeAs within spin-fluctuation theory\cite{Wang13} (see Ref. \onlinecite{HirschfeldCRAS} and references therein).  
      
    The consequences of correlations for the band structure of FeSC are more profound than simple Fermi surface shifts, however.   If one examines compounds where the $d$-bands are closer to half-filling (5 electrons/Fe), the effect of  electron-electron interactions are enhanced in a way distinctly different from  one-band systems: different
      $d$  orbital effective masses are enhanced by different factors.  This ``orbital selectivity'' predicted by theory\cite{deMedici_review,Bascones_review,Biermann_review,Yu11, Yi13, Yu13} has been confirmed by ARPES experiments.  While most Fe-based systems have more electrons/Fe, closer to 6, the effects are still nontrivial in the Fe-chalcogenides.  For example, the electrons in bands with  $d_{xy}$ orbital character have been claimed to exhibit single particle masses up to 10-20 times the band mass, while in $d_{xz}/d_{zy}$ states the renormalization is closer to 3-4\cite{Yi2015,Liu2015}. 

In Fermi liquid theory, excitations in a system of interacting fermions are described by quasiparticles that have the same quantum numbers but deviate from the free particles in properties such as the quasiparticle mass,  which renormalizes the Fermi velocity.  Generally, interactions in electronic systems also lead to  reduced quasiparticle weights, corresponding to  reduced values of  the residue at the pole of the Green's function describing those dressed electrons.
      Even in one-band systems where orbital selectivity does not play a role, pairing in superfluid systems with reduced Landau quasiparticle weight is an important unsolved theoretical problem.  While one 
      generally expects pairing interactions to be reduced as the quasiparticle weight is suppressed as other aspects of pairing are held fixed,  pairing in completely
      incoherent non-Fermi liquids is not impossible, as discussed recently in Ref. \onlinecite{Sachdev_incoherent}.    The effect of  orbital selective quasiparticle weights on pairing in FeSC has been discussed elsewhere in various approximations\cite{Ogata_selectivepairing,Si_selectivepairing}, with differing conclusions.

      In this work, we  implement a simple scheme to incorporate aspects of renormalization of the electronic band structure, including reduced quasiparticle coherence that is orbital selective into spin-fluctuation pairing theory, and apply it to several FeSC. This orbital selective approach to pairing provides an excellent description for the superconducting gap deduced from quasiparticle interference measurements on the nematic
      Fermi surface pockets of bulk FeSe, as shown already in Ref. \onlinecite{Davis_OSP}.  Here we discuss the generality of this approach, and
      show how it explains the exotic gap structures of FeSe, 
      FeSe monolayers 
      and in the LiFeAs system as well.
      These findings encourage us to believe that the proposed paradigm is the correct way to understand the physics in these materials, but we cannot rule out completely that other effects affecting the gap such as spin-orbit coupling or orbital fluctuations\cite{Saito14} may contribute.
      While the microscopic origin of the phenomenology remains an open challenge, we believe that it provides a major step towards a quantitative, material-specific  theory of superconductivity in strongly correlated FeSC.\\

{\ital{Model}}
The starting point of any uncorrelated multiband system is the electronic structure described by a tight-binding model\cite{Eschrig09,Wang13,Ahn14,Saito14}
\begin{equation}
H=\sum_{\mathbf{k}\sigma \ell \ell'} t^{\ell \ell'}_{\mathbf{k}} c_{\ell\sigma}^\dagger(\k) c_{\ell'\sigma}(\k),
 \label{eq_tb}
\end{equation}
where $c_{\ell\sigma}^\dagger(\k)$ is the Fourier amplitude of an operator that creates an electron in  Wannier orbital $\ell$ with spin $\sigma$ and $t^{\ell \ell'}_{\mathbf{k}}$ is the Fourier transform of the hoppings.
By a unitary transformation from orbital to band space, $H$ becomes diagonal \mbox{$H=\sum_{\mathbf{k}\sigma \mu} \xi_\mu(\k)c_{\mu\sigma}^\dagger(\k)c_{\mu\sigma}(\k)$}, with eigenenergies $\xi_\mu(\k)$ and $c_{\mu\sigma}^\dagger(\k)$ creating an electron in Bloch state $\mu,\k$.

There is no way to determine empirically the electronic structure $\xi_\mu(\k)$ of the uncorrelated reference system corresponding to a given real material.  However, experimental probes like ARPES and quantum oscillations provide information on the real single-particle spectrum, which we will call $\tilde E_\mu(\k)$.    Since we do not have access to $\xi_\mu(\k)$,  we will henceforth use the term ``uncorrelated'' to mean a model for an electronic structure where the quasiparticles have unit weight; in this work we only work with such models where the eigenenergies $\tilde E_\mu(\k)$ have been obtained by fit to experiment.
In Fig. \ref{fig_fermi} we show examples of Fermi surfaces derived from the eigenenergies $\tilde E_\mu(\k)$. For three dimensional (3D) models considered in this work, the zero energy surfaces, i.e. the set of $\k$ vectors with  $\tilde E_\mu(\k)=0$ are  corrugated tubes identified as $\alpha$, $\delta$ and $\varepsilon$ sheets in Fig. \ref{fig_fermi}(a) (FeSe, bulk) or the $\beta$ and $\gamma$ sheets in (c) (LiFeAs), but can also be closed surfaces
as the $\alpha$ pocket in (c).  For a 2D model as shown in (b), the Fermi surface is given by elliptical lines such that it is convenient to plot quantities as a function of the angle $\varphi$.
\begin{figure}[tb]
\includegraphics[width=\linewidth]{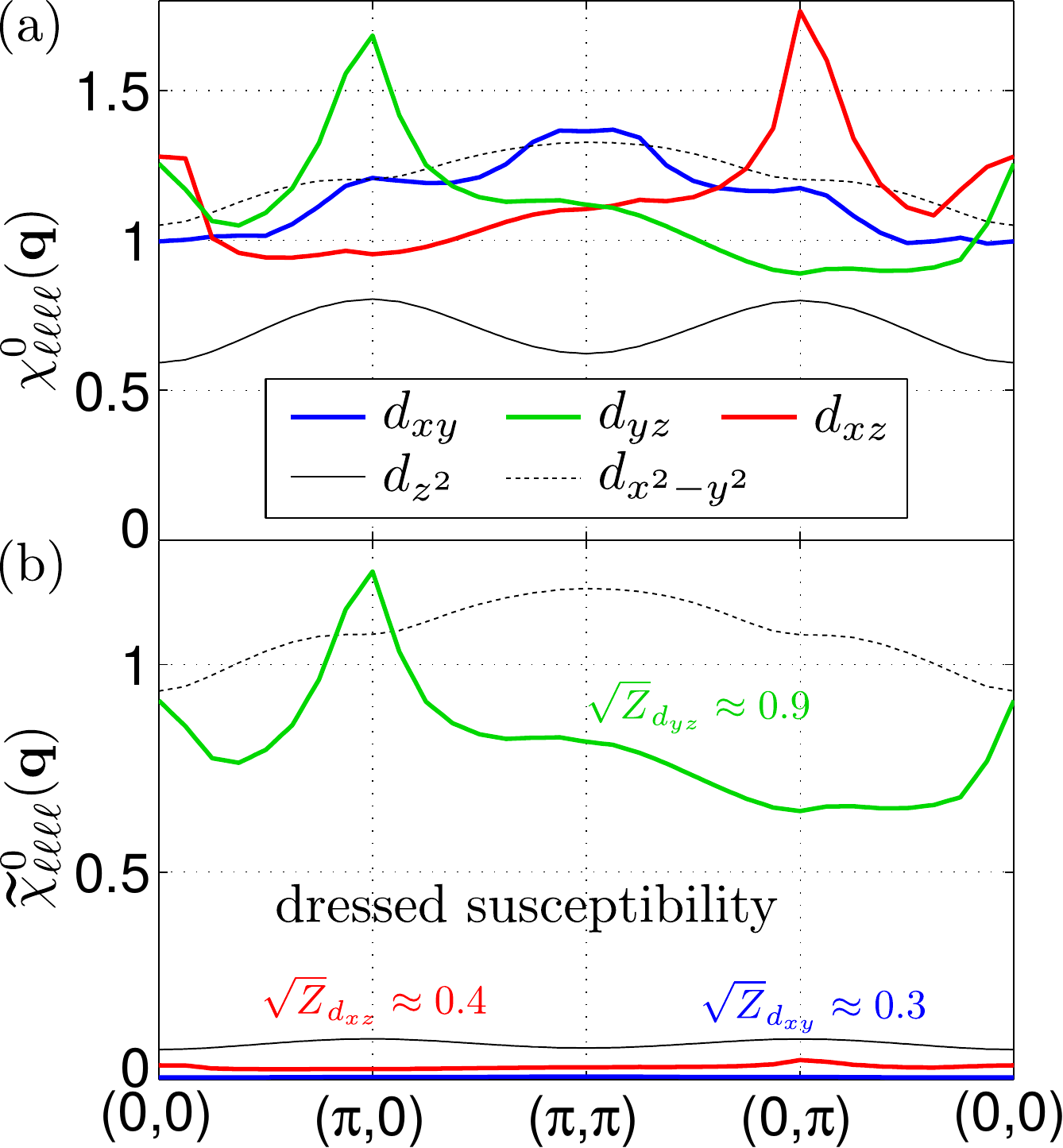}
\caption{Comparison of the orbitally diagonal components of the susceptibility of the uncorrelated model for bulk FeSe (a) and the same quantities including the quasiparticle weights that suppress contributions from orbitals with small weight factors according to Eq. (\ref{eq_susc}) (b).
\label{fig_susc}}
\end{figure}

In the orbital basis the  ``uncorrelated" Green's function is given by
\begin{align}\label{eq_Gorb}
G_{\ell\ell'} (\k,\omega_n)= \sum_\mu \frac{a_\mu^\ell(\k) a_\mu^{\ell'*}(\k)}{i \omega_n -  \tilde E_\mu(\k)},
\end{align}
where $a_\mu^\ell(\k)$ are the matrix elements of the unitary transformation mentioned above. 
The orbital weight $|a_\mu^\ell(\k)|^2$ becomes important when discussing low-energy (Fermi-surface driven) properties and is therefore visualized color coded for the important Fe $d$ orbitals
$\ell=\{d_{xy},d_{xz},d_{yz}\}$ in Fig. \ref{fig_fermi} as well.

In order to include the full effects of correlations, we further make the orbital selective ansatz that the operators $c^\dagger_\ell (\k)$ create quasiparticles with weight $\sqrt{Z_\ell}$ in orbital $\ell$,  $
  c^\dagger_\ell(\k) \rightarrow\sqrt{Z_\ell} c^\dagger_\ell (\k).$  Note that $\ell$ runs over the Fe $3d$ orbitals  $(d_{xy},d_{x^2-y^2},d_{xz},d_{yz},d_{3z^2-r^2})$.
\begin{figure}[b]
\includegraphics[width=\linewidth]{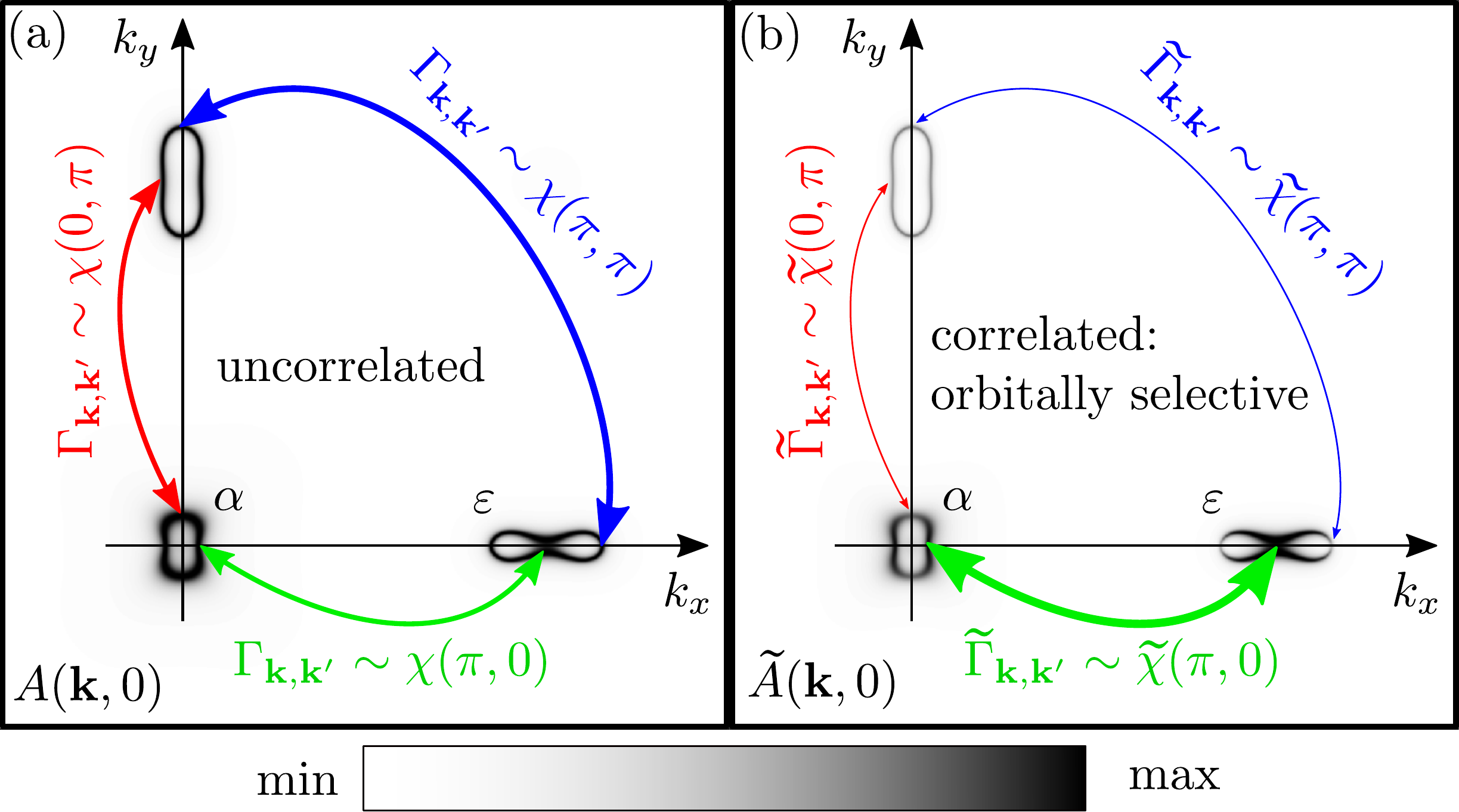}
\caption{ Plot of the spectral function at zero energy in the first Brillouin zone. (a) Spectral function $A(\k,0)=-1/\pi \mathop{\text{Im}} \mathop{\text{Tr}} G(\k,0)$ of the uncorrelated model for FeSe (bulk) at $k_z=0$ with the Green's function as in Eq. (\ref{eq_Gorb}). (b) Spectral function $\tilde A(\k,0)$ of the model including quasiparticle weights inducing orbital selective reduced coherence. For the pair scattering of Cooper pairs at momenta $\k$ to $\k'$ on the Fermi surface (arrows) two quantities determine the scattering strength: (i) the susceptibility $\tilde \chi(\q)$ to which the pairing vertex $\Gamma_{\k,\k'}$ is proportional and (ii) the quasiparticle weight at initial and final  momentum. In summary, some processes get largely suppressed (thin red and blue arrows) such that other processes (thick green arrow) dominate the Cooper pairing.
\label{fig_pair}}
\end{figure}
\begin{figure*}[tb]
\includegraphics[width=\linewidth]{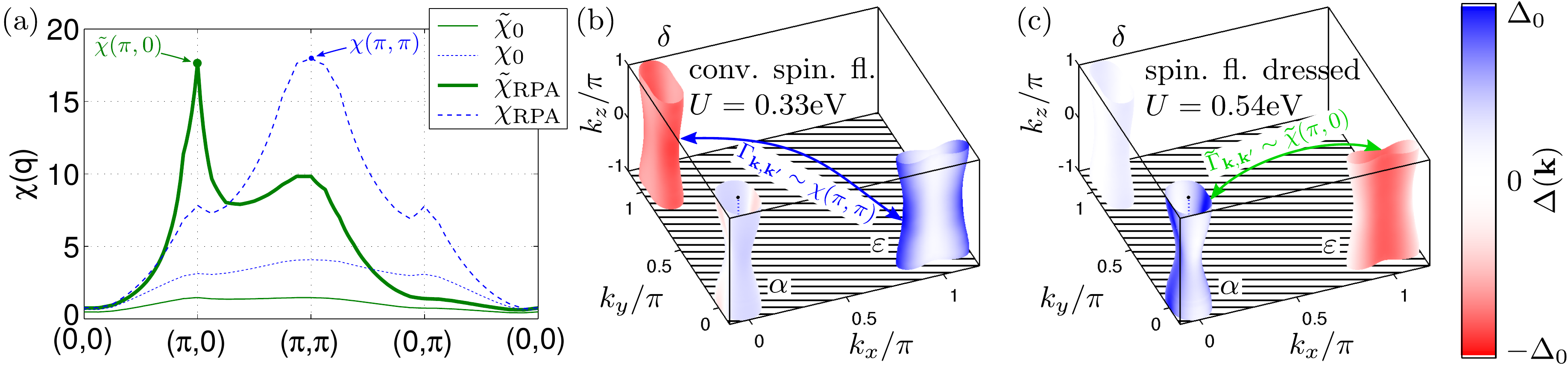}
\caption{ Results for FeSe (bulk): (a) Calculated susceptibility with quasiparticle weights ($ \tilde \chi$, thick lines) compared to the susceptibility without quasiparticle weights ($\chi$, thin dashed lines), (b) gap symmetry function as obtained from conventional spin-fluctuation pairing and (c) the same quantity when taking into account orbital dependent quasiparticle weights. For both calculations, the dominant pair scattering processes leading to a large order parameter are symbolized with a double arrow. The calculations are done for a fixed ratio $J=U/6$, but with an overall scale $U$ as indicated.}
\label{Fig_FeSe_bulk}
\end{figure*}
The associated Green's function becomes
\begin{equation}\label{eq:5}
\tilde G_{\ell\ell'} (\k,\omega_n) =\sqrt{Z_\ell  Z_{\ell'}} \;\sum_\mu \frac{a_\mu^\ell(\k) a_\mu^{\ell'*}(\k)}{i \omega_n - \tilde E_\mu(\k)},
\end{equation}
where $\tilde E_\mu(\k)$ are the renormalized band energies.
A similar approach has been used recently when parametrizing the normal state Green's functions in a Fermi liquid picture\cite{Yamakawa_PRX_16}, with the formal difference that we explicitly employ the renormalized quasiparticle energies $\tilde E_\mu(\k)$, which include the static real
part of the self-energy, and retain the quasiparticle weights in the numerator.
Following state-of-the-art pairing calculations from spin-fluctuation theory\cite{s_graser_09,a_kemper_10,Kreisel13,Wang13} (see Appendix \ref{ap_sp_fl}),
important effects of the $\sqrt{Z_\ell}$ factors enter in two places: 1) the calculation of the susceptibility includes the renormalized quasiparticle Green's function, and 2) when projecting the pairing interaction from orbital to band space, one needs to account for the replacement of $
  c^\dagger_\ell(\k) \rightarrow\sqrt{Z_\ell} c^\dagger_\ell (\k)$. In cases where the Hamiltonian already correctly describes the quasiparticle energies of a correlated system $\xi_\mu(\k) \rightarrow \tilde E_\mu(\k)$ (as obtained, e.g., from fits to measured quasiparticle energies from spectroscopic experiments), the bare susceptibility in orbital space needs to be simply multiplied by the quasiparticle weights
\begin{equation}
 \tilde \chi_{\ell_1 \ell_2 \ell_3 \ell_4}^0 (\q)=\sqrt{Z_{\ell_1}Z_{\ell_2}Z_{\ell_3}Z_{\ell_4}}\;\chi_{\ell_1 \ell_2 \ell_3 \ell_4}^0 (\q),
 \label{eq_susc}
\end{equation}
in order to obtain the corresponding quantity (with tilde) in the correlated system.
Our models as shown in \mbox{Fig. \ref{fig_fermi}} already match the true quasiparticle energies $\tilde E_\mu(\k)$, such that we can use Eq. (\ref{eq_susc}) to examine the effect of the quasiparticle weights on the susceptibility. In Fig. \ref{fig_susc}(a),  the diagonal components of the orbitally resolved susceptibilities where $\ell_1= \ell_2 =\ell_3= \ell_4$ are plotted as obtained from our model of FeSe (bulk). For all orbitals, the overall magnitude is similar (except for $\ell=d_{z^2}$ that does not play any role for the subsequent discussion), but the momentum structure is distinct: The $d_{xy}$ component has a maximum at $\q=(\pi,\pi)$, whereas the components for $d_{yz}$ ($d_{xz}$) have maxima at $\q=(\pi,0)$ ($\q=(0,\pi)$). Introducing quasiparticle weights as indicated in \mbox{Fig. \ref{fig_susc}(b)}, it is obvious that some components are suppressed more than others such that for the present choice of $\{\sqrt{Z_l}\}=[0.2715,0.9717,0.4048,0.9236,0.5916]$, the $d_{yz}$ contribution dominates\footnote{Note that the $d_{x^2-y^2}$ component is still large because of the choice of a quasiparticle weight close to 1. It therefore contributes to the physical susceptibility, but has little influence on the supercondicting order parameter since the orbital weight for states at low energies is small, see Fig. \ref{fig_fermi}}.
In a similar way, the pairing interaction gets modified by prefactors from quasiparticle weights (see Appendix \ref{ap_sp_fl}). Physically, this means  that orbital-selective pairing occurs because pairing from certain quasiparticle states is suppressed more than others because the states themselves are less coherent.

To visualize this effect, we have plotted the spectral function $A(\k,\omega)=-1/\pi \mathop{\text{Im}} \mathop{\text{Tr}} G(\k,\omega)$ for $k_z=0$ at zero energy in Fig. \ref{fig_pair}(a) for the uncorrelated system and in (b) with the same choice of quasiparticle weights as discussed above. We use the bulk FeSe Fermi surface discussed below as an illustration of the idea, but details of the bands are not important for this purpose. The superconducting order parameter is now determined by the strength of the pair scattering $\Gamma_{\k,\k'}$ of a Cooper pair at $\k$ to $\k'$ which is proportional to the susceptibility within the spin-fluctuation approach. In the uncorrelated case, scattering processes involving three pairs of $\k$-vectors as depicted by the arrows in Fig.~\ref{fig_pair} are comparable in magnitude (with the process in blue involving $d_{xy}$ states being slightly larger). Taking into account the quasiparticle weights, the spectral function and thus the pair scattering is suppressed on parts of the Fermi surface. Consequently, the processes involving $d_{yz}$ states (green, thick arrow) dominate over those involving $d_{xy}$ states (blue) and $d_{xz}$ states (red), making the pairing orbital selective.
\\

{\ital{Bulk \texorpdfstring{F\MakeLowercase{e}S\MakeLowercase{e}}{FeSe}}}  Early thermodynamic and transport studies of bulk FeSe, as well as STM supported a state with gap nodes\cite{Song2011,Kasahara2014}.
 However, more recent measurements of 
low-temperature specific heat \cite{Lin2011,Jiao2016},  STM \cite{Jiao2016},   thermal
conductivity\cite{Dong2009,Bourgeois2016} and penetration depth\cite{Li2016,Teknowijoyo2016} have found a tiny spectral gap, indicating that the gap function is highly anisotropic but may not change sign on any given sheet. The only experiments that provide information on the location of these deep minima are an ARPES measurement on the related Fe(Se,S) material\cite{Xu2016} and a recent quasiparticle interference (QPI) experiment\cite{Davis_OSP}, both of which find deep minima on the tips of the hole ellipse  at the center of the Brillouin zone. 
The latter also distinguishes deep minima on the tips of the $\varepsilon$ electron pocket ``ellipse''.

To test the mechanism of orbital selective pairing determined by reduced coherence of some quasiparticles, we show first how this mechanism modifies results for the susceptibility and the superconducting gap for bulk FeSe. Our starting point is a tight-binding model with hoppings adapted such that the spectral positions of the quasiparticle energies fit recent findings using ARPES, quantum oscillations and STM experiments\cite{Terashima2014,Audouard2015EPL_Hc2,Watson15,Watson2016,Davis_OSP}.  As the band energies are ``measured'' in this case, these can be identified with the renormalized band energies $\tilde E_\mu(\k)$ in the presence of correlations, yielding the Fermi surface in Fig. \ref{fig_fermi}(a).

To construct a proper approximation of the quasiparticle Green's function [Eq. (\ref{eq:5})], we need to additionally include quasiparticle weights. Next, we fix the ratio $J=U/6$ as found in cRPA calculations\cite{Miyake_cRPA,Scherer_FeSe} and optimize the weights in the orbital basis.
The result is $\{\sqrt{Z_l}\}=[0.2715,0.9717,0.4048,0.9236,0.5916]$ such that the gap function yields a nodeless order parameter with a large anisotropic gap on the $\alpha$ pocket, as seen from Fig. \ref{Fig_FeSe_bulk}(c).
These values for $Z_l$ are in reasonable agreement with general trends in FeSC: the $d_{xy}$ orbital exhibits strongest correlations (smallest weight)\cite{Yi2015}, while the $d_{x^2-y^2}$ orbital is the most weakly correlated\cite{deMedici_review,Bascones_review,Biermann_review}.
We note that the resulting gap structure is very different from the one obtained from conventional spin-fluctuation calculations (which also show a distortion from tetragonal symmetry as expected)\cite{Kreisel15}, a result of the very different momentum structure of the pairing interaction [compare Fig. \ref{Fig_FeSe_bulk}(b,c)]: The largest gap magnitude is on the tip electron pocket ($\varepsilon$) centered at the X point for the conventional calculation, because the largest pair scattering $\Gamma_{\k,\k'}$ connects this area of the Fermi surface with the corresponding one on the Y centered pocket [blue arrow in Figs. \ref{fig_pair}(a) and \ref{Fig_FeSe_bulk}(b)]. It appears on the $\alpha$ pocket when using the orbital selective pairing ansatz, because the dressed electrons mediate the strongest Cooper pair scattering from the flat area of the $\alpha$ pocket to the flat area of the $\varepsilon$ pocket, where also the gap is maximal [green arrow in Figs. \ref{fig_pair}(b) and \ref{Fig_FeSe_bulk}(c)].
The physical origin of this can be attributed to the strong splitting of weights of the $d_{xz}$ and $d_{yz}$ orbitals where states of the $d_{xz}$ orbital are very incoherent.

\begin{figure}[tb]
\includegraphics[width=\linewidth]{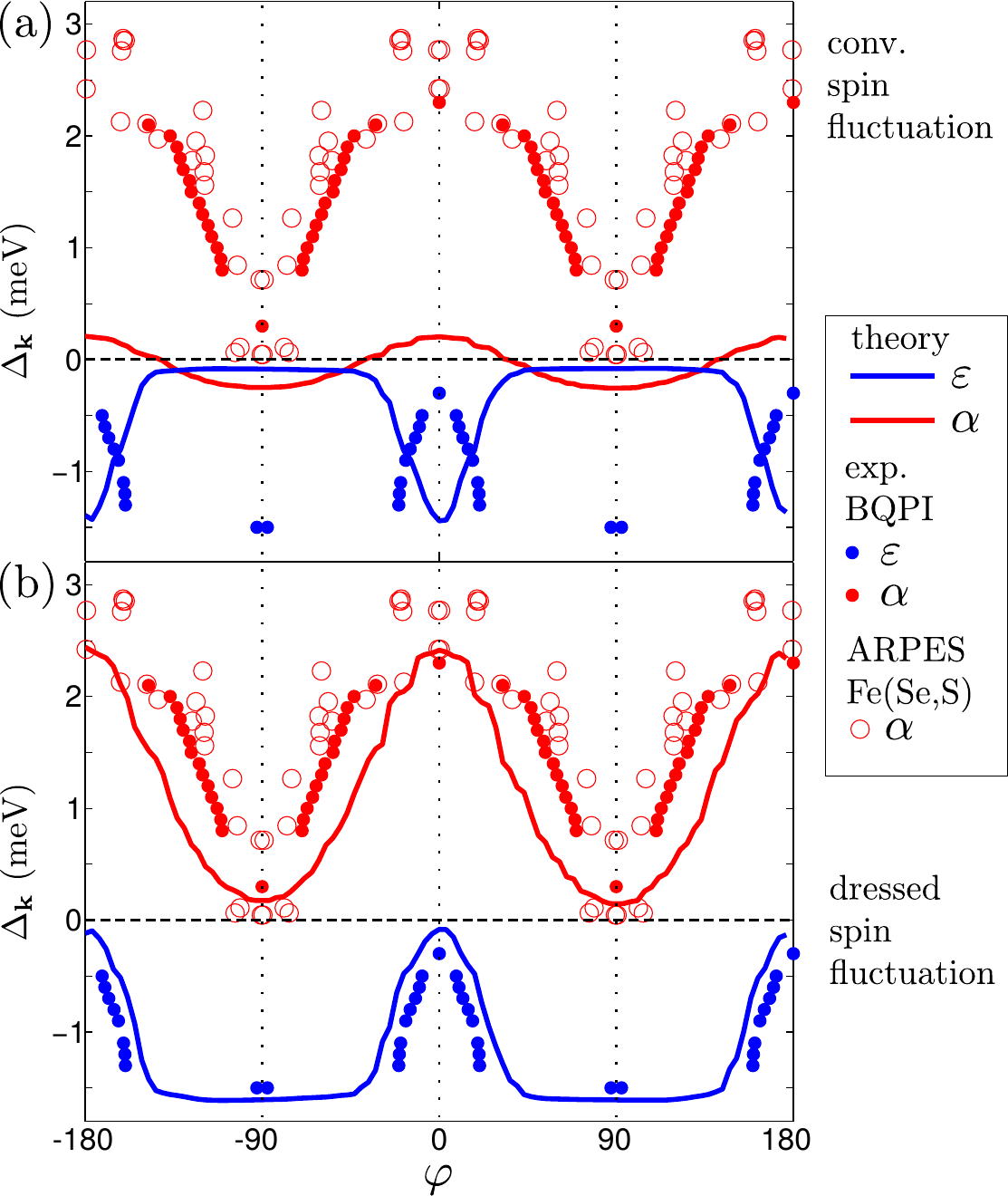}
\caption{ Results for FeSe (bulk): Plot of the gap function around the Fermi surface pockets for (a) the conventional spin fluctuation calculation and (b) a calculation using the spin-fluctuation pairing in presence of quasiparticle weights. For direct comparison, the data from a Bogoliubov QPI analysis from Ref.\cite{Davis_OSP} and a ARPES investigation on a related compound FeSe(S)\cite{Xu2016} are displayed as well.
\label{fig_FeSe_cut}}
\end{figure}
We observe that the susceptibility $\tilde \chi$, originally strongly dominated by $(\pi,\pi)$, now shows dominant stripe fluctuations with $\q=(\pi,0)$ [see Fig. \ref{Fig_FeSe_bulk}(a)].
This result is in agreement with findings from neutron scattering experiments\cite{Wang2015,WangZhao2016NatMat_FeSe-nematicity} which find strong stripe fluctuations at low energies. Taking into account the results of a recent ARPES experiment\cite{Kushnirenko2017} with the conclusion that the electronic structure of FeSe evolves in such a way that it becomes less correlated as temperature increases, we can conclude that weight of the spin-fluctuations should shift from $(\pi,0)$ towards $(\pi,\pi)$ as temperature increases.
This can be understood directly from Eq. (\ref{eq_susc}), where the different orbital components of the susceptibility are weighted according to the quasiparticle weights; the $d_{xy}$ components which are peaked at $(\pi,\pi)$ get suppressed. The $d_{xz}$ components, peaked at $(0,\pi)$, are
suppressed as well (see Fig. \ref{fig_susc}). On individual pockets, the gap function then follows the orbital content of the orbital with strongest contribution (in this case, the $d_{yz}$ orbital) [compare Fig. \ref{fig_fermi} (a)].

Consequently, the pairing is changed by two mechanisms: First, it is modified directly  by the quasiparticle weights as discussed earlier and, second, the peak shifts in $\q$ in the (RPA) susceptibility. Both of these effects make the pair scattering in the $d_{yz}$ orbital more important [green thick arrow in Fig. \ref{fig_pair}(b)] yielding the gap structure as shown in Fig. \ref{Fig_FeSe_bulk}(c). To make the agreement to experiment evident, we plot in Fig. \ref{fig_FeSe_cut} the gap function at a cut of the Fermi surface at $k_z=\pi$ comparing to results from two different spectroscopic methods. While the conventional calculation [\ref{fig_FeSe_cut} (a)] does not show any similarities, the correspondence in (b) is evident. Finally, we 
note that this picture is different than that ascribed to orbital selective physics in the ``strong-coupling"
$t-J$ model approach, where the $d_{xy}$ pairing channel is enhanced rather than suppressed\cite{Si_selectivepairing}.\\

\begin{figure*}[bt]
\includegraphics[width=\linewidth]{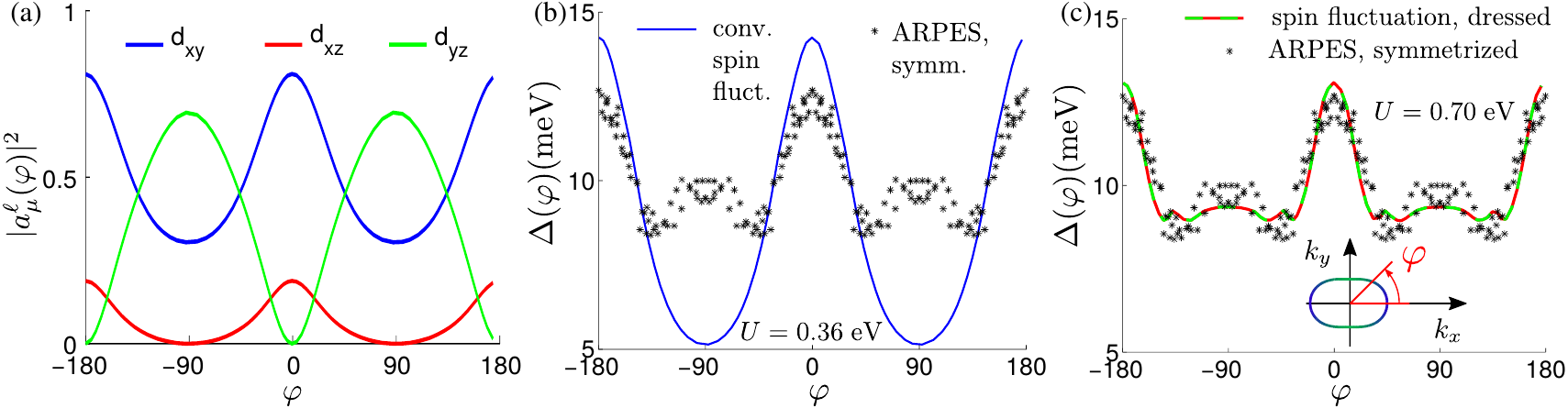}
\caption{ Results for monolayer FeSe: (a) Orbital weight at the Fermi surface. (b) Superconducting gap obtained from conventional spin-fluctuation theory, and (c) the same quantity including orbital dependent quasiparticle weights compared to measured gap functions in ARPES.\cite{Zhang16} Symmetry operations of the tetragonal system have been applied to the measured data.
All calculations were done for a fixed ratio $J=U/10$, with overall scale $U$ as indicated.}
\label{Fig_FeSe_mono}
\end{figure*}

{\ital{Monolayer \texorpdfstring{F\MakeLowercase{e}S\MakeLowercase{e}}{FeSe} on \texorpdfstring{S\MakeLowercase{r}T\MakeLowercase{i}O$_3$}{SrTiO$_3$}}} Despite considerable excitement over the high critical temperature in the FeSe/STO monolayer system, limited information is available regarding the structure of the superconducting gap.  Early ARPES measurements suggested an isotropic gap on electron pockets\cite{Liu2012,Tan13}.   Theoretical possibilities for pairing states in the presence of missing $\Gamma$-centered hole band were discussed in Ref. \onlinecite{HirschfeldCRAS}.  Quite recently, a new ARPES study identified significant and unusual anisotropy on a single unhybridized elliptical electron pocket\cite{Zhang16}, whereby the gap acquired global maxima at the ellipse tips, and additional local maxima at the ellipse sides.  These authors showed that the structure cannot be explained using any of the low-order  Brillouin zone harmonics expected from so-called ``strong coupling'' electronic pairing  theories.

Within the model for the electronic structure of bulk FeSe, we perform a  calculation with a few modifications to account for differences in the monolayer from the bulk:
(1) We ignore all hoppings out of the plane, yielding a strictly 2D system.
(2) We neglect orbital order, which has never been observed in the monolayer.
(3) Experimentally, only electron-like Fermi pockets have been detected, suggesting that the monolayer is actually electron doped. Possible reasons for this doping are charge transfers from the substrate or surface defects.
We therefore apply a rigid band
shift by $\delta\mu=60 \;\text{meV}$, which removes the $\Gamma$-centered hole pocket and leaves electron pockets that have
the size and shape of measured spectral functions in ARPES\cite{Zhang16}, with $n=6.12$ electrons/Fe, see \mbox{Figs. \ref{fig_fermi}(b)} and \ref{Fig_FeSe_mono}(a) for a plot of the orbital character. The quasiparticle weights in the monolayer may be different from the bulk for two reasons: (1) The absence of the orbital order, i.e., the tetragonal crystal structure dictates that the weights for  $d_{xz}$ and $d_{yz}$ orbitals are degenerate (unlike bulk FeSe). (2) Correlations may be different in the monolayer where a tendency towards weaker correlations was found recently\cite{Haule16}, such that we fix the ratio $J=U/10$ in this case.

At this point, we note that the states on the Fermi surface have only tiny orbital weight of $d_{z^2}$ and $d_{x^2-y^2}$ character, and additionally there are no pair scattering processes from $\k$ to $\k'$ with $\q=(\pi,0)$ [or  $\q=(0,\pi)$] such that a fit procedure with all quasiparticle weights will be under-determined. In the optimization procedure, we therefore fix the weights to $\sqrt{Z_{x^2-y^2}}=0.8 > \sqrt{Z_{z^2}}=0.7$ and obtain
$\{\sqrt{Z_l}\}=[ 0.4273 ,   0.8000  ,  0.9826 ,   0.9826 , 0.700]$ for the best agreement to the gap measured in ARPES\cite{Zhang16}.
This result  does change the susceptibility slightly, but keeps the $(\pi,\pi)$ fluctuations dominant; for details we refer to Fig. \ref{Chi_monolayer_FeSe} in the Appendix. These fluctuations drive an overall (nodeless) $d$-symmetry ground state as expected, but with an unusual structure modified strongly by orbital correlations,
 with the result as shown in Fig. \ref{Fig_FeSe_mono}(b,c).
Evidently the gap function for the standard spin-fluctuation calculation [Fig.~\ref{Fig_FeSe_mono}(b)] mostly follows the orbital content of the $d_{xy}$ orbital [compare Fig. \ref{Fig_FeSe_mono}(a) for a plot of the orbital weights as a function of angle $\varphi$ around the X-centered pocket\footnote{The Y-centered pocket is symmetry related and will not be discussed further at this point}]. For the current Fermi surface, this is expected because the pairing interaction is dominated by intra-orbital processes, and the $d_{xy}$ orbital has large weight at positions $\k$ and $\k'$ on the Fermi surface which are separated roughly by $(\pi,\pi)$ and can take advantage of the strong peak in the susceptibility at that $\q$ vector. The other two orbitals play a negligible role in the pairing process. This situation is modified once the pairing interaction is renormalized by the quasiparticle weights and therefore reduces the contribution of the $d_{xy}$ orbital.  The main effect is that a second maximum in the gap function appears at a position in momentum space where the $d_{xz}$ or $d_{yz}$ orbital is dominant [see Fig. \ref{Fig_FeSe_mono} (c)].

\begin{figure*}[tb]
\includegraphics[width=\linewidth]{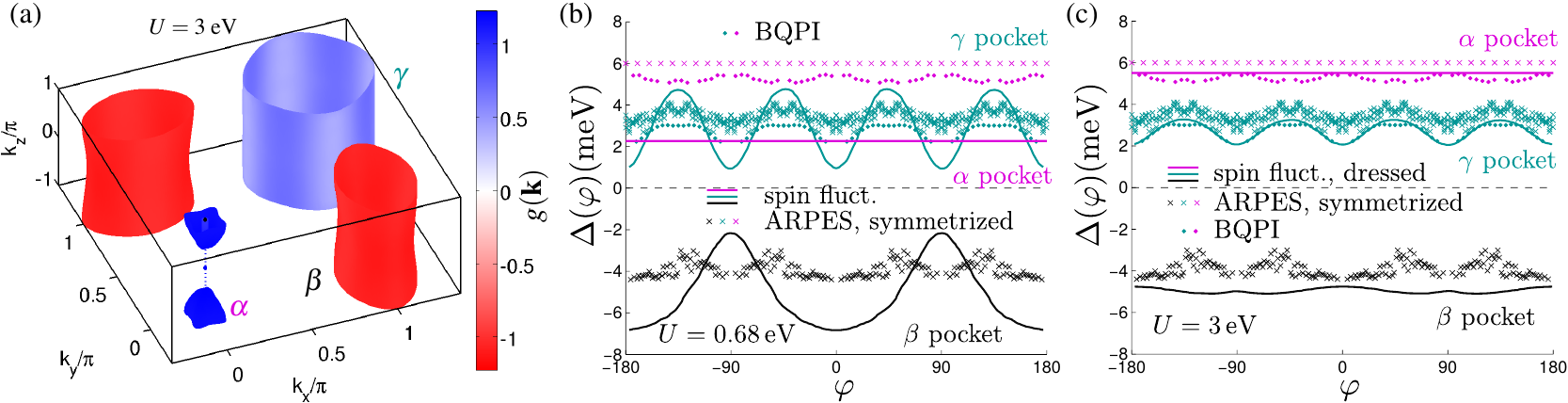}
\caption{ Results for LiFeAs: (a) 3D plot of the gap function as obtained from spin-fluctuation calculation including quasiparticle weights. (b) Cut at $k_z=\pi$ of the result of the s-wave gap function from conventional spin-fluctuation theory (solid lines) plotted as a function of angle $\varphi$ (as defined in Fig. \ref{fig_fermi}) around the pockets ($\Gamma$-centered hole pocket ($\alpha$, magenta), M-centered hole pocket ($\gamma$, cyan) and X-centered electron pocket ($\beta$, black)) together with experimental results. The measured magnitudes of the gap from an ARPES experiment\cite{Borisenko12} are symmetrized and displayed as crosses, and those from a Bogoliubov QPI experiment\cite{Allan12} as filled dots. (c) The same quantity for the gap function as shown in (a) also compared to experimental data. All calculations are done for a fixed ratio $J=0.37U$\cite{Wang13}, but with overall scale $U$ as indicated.
\label{Fig_LiFeAs}}
\end{figure*}

In the pairing process,  intra-orbital, {\it inter-pocket} contributions dominate, whereby one pair on the $X$ pocket of $d_{yz}$ character  scatters into another pair on the $Y$ pocket with the same orbital character, meaning that the latter pair  must be located on the tip of the $Y$-pocket where the gap has largest magnitude. Because the total weight of this orbital is smaller there, the order parameter for $\k$ states dominated by this orbital is enhanced. In summary, one  gets a gap structure with a large maximum at the tip of the ellipse and a small maximum at the flat part of the ellipse, remarkably similar to that detected by experiment.\cite{Zhang16}\\

{\ital{\texorpdfstring{L\MakeLowercase{i}F\MakeLowercase{e}A\MakeLowercase{s}}{LiFeAs}}} LiFeAs is another Fe-based superconductor that is known to have a Fermi surface quite different from that predicted from DFT.  Several theoretical attempts\cite{Wang13,Ahn14,Saito14,Yin2014} to understand the ARPES-determined gap structure\cite{BorisenkoLiFeAs,Borisenko12,Umezawa12,Allan12} were reviewed recently in Ref. \onlinecite{HirschfeldCRAS}.  All were based 
on an ``engineered'' tight-binding band structure consistent with ARPES data\cite{Wang13}, i.e., containing the correct spectral positions of the bands (including the orbital content).
Despite some success in explaining certain features of the gap structure, others were not reproduced properly in all approaches, although Ref. \onlinecite{Saito14} claimed a good overall fit to experiment.

To reveal how and whether the standard spin-fluctuation theory result changes upon inclusion of quasiparticle weights, we use the same method as described above for a band structure relevant to  LiFeAs\cite{Wang13}.  The corresponding Fermi surface is shown in Fig. \ref{fig_fermi}(c).
First, we note that moderate changes in the quasiparticle weights which we set to $\{\sqrt{Z_l}\}=[0.5493,0.969,0.5952,0.5952,0.9267]$  do change the gap structure, but largely preserve the structure of the susceptibility (see Appendix \ref{sp_fl_qp}).
The gap functions, however,  undergo a remarkable change relative to unrenormalized spin-fluctuation theory.  These include first
 a stronger tendency towards $s_\pm$ symmetry, even with small values of $J$. Note that the conventional spin-fluctuation scenario, $d$ and $s$ wave solutions are nearly degenerate, a consequence of the poor $(\pi,0)$ nesting properties of LiFeAs\cite{BorisenkoLiFeAs,Lee_etal_Kotliar2012}.
Secondly, orbital selectivity enhances the gap on the small $\Gamma$-centered hole pocket ($\alpha$ pocket), see Fig. \ref{Fig_LiFeAs}(a).  This appears to correct the crucial discrepancy in  the calculation of Wang {\it et al}.\cite{Wang13} relative to experiment [see Fig. \ref{Fig_LiFeAs} (b,c)].  Finally, the procedure leads to weaker anisotropy of the gap on the large $d_{xy}$ dominated pocket, also in better agreement with experiment, whereas small deviations between the ARPES data\cite{BorisenkoLiFeAs} and our calculation on the electron pockets persist which could be due to hybridization of the corresponding bands. We did not investigate effects of spin-orbit coupling in this case since these are supposed to be small\cite{Wang13}. Note further that the (angular) position of the maximum gap on the electron pockets change from 0 degrees to slightly off 90 degrees, opening the possibility of two maxima (and two minima). Unlike the models for FeSe (bulk) and monolayer FeSe, all three orbitals ($d_{xy}$, $d_{xz}$, $d_{yz}$) play an important role in determining the gap anisotropy on the $\beta$ pockets, making it more sensitive to changes in the electronic structure.\\

{\ital{Discussion}} The above results are extremely encouraging,
suggesting that the orbital selective
correlation effects are indeed 
required when applying
spin-fluctuation pairing theory to Fe-chalcogenide
and more strongly correlated Fe-based superconductors.  
We caution, however, that we have not derived the renormalizations
entering the pair vertex self-consistently from a microscopic theory.
Efforts along these lines are  in progress.   Secondly, by construction
 the quasiparticle renormalizations
describe only the states near the Fermi level.  Comparison with ARPES measurements should be performed carefully, as these analyses tend to emphasize  renormalizations on much larger energy scales, which may be quite different.
Possible imprints of the orbital selectivity could be visible in the penetration depth\cite{Li2016} if calculated within the same theoretical framework, or Friedel oscillations close to impurities in the case of bulk FeSe which are rotating in direction as a function of energy\cite{Kasahara2014}. Calculations along these lines are also in progress.
\\

{\ital{Conclusions}} 
In the absence of a fully controlled many-body treatment of electronically paired superconductivity, it may be very valuable to have a simple phenomenological yet microscopic approach that includes aspects of the low-energy quasiparticle renormalizations that affect
pairing most strongly.  We have presented a paradigm  that allows for suppressed quasiparticle weight within the framework of conventional spin-fluctuation pairing theory, and argued that it
provides accurate descriptions for the
previously inexplicable superconducting energy gap structures of the most strongly correlated FeSC.
We have given  results of explicit calculations in three cases where correlations are known to play an important role, bulk FeSe, monolayer FeSe on STO, and LiFeAs.
These results reveal an immediate challenge to determine
if our approach can be combined with microscopic calculations of quasiparticle weights to yield a material-specific theory with predictive power for strongly correlated FeSC.\\

\begin{acknowledgments}
We would like to thank A.V. Chubukov, D.J. Scalapino, and D. D. Scherer for useful discussions.   A.Kr. and B.M.A. acknowledge support from a Lundbeckfond fellowship (Grant No. A9318). P.J.H. acknowledges support  through Department of Energy DE-FG02-05ER46236.
J.C.S.D. acknowledges gratefully support from the Moore Foundation’s EPiQS Initiative through Grant GBMF4544, and the hospitality and support of the Tyndall National Institute, University College Cork, Cork, Ireland. P.O.S. and A.Ko. acknowledge support from the Center for Emergent Superconductivity, an Energy Frontier Research Center, headquartered at Brookhaven National Laboratory and funded by the U.S. Department of Energy under DE-2009-BNL-PM015.
  \end{acknowledgments}
\appendix
\renewcommand\thefigure{S \arabic{figure}}    
\setcounter{figure}{0}
\section{Hamiltonian and construction of Green's function}
Considering the tight binding Hamiltonian, Eq. (\ref{eq_tb}) together with its diagonalization to band basis, one can construct the Green's function in the band basis $G_\mu(\k, \omega_n) = [i\omega_n-\xi_\mu(\k)]^{-1}$. The
unitary transformation that takes one from the band basis (Greek indices) to the orbital basis (Roman indices) is
  \begin{equation}
  c_{\ell\sigma}(\k)=\sum_{\nu}a^\ell_\nu(\k) c_{\nu\sigma}(\k).
\label{eq:1}
\end{equation}  
Unitarity implies
\begin{equation}\label{eq:2}
\sum_\ell  a^\ell_\nu(\k) a^\ell_\mu(\k)^* = \delta_{\mu\nu}
\end{equation}
so we can invert (\ref{eq:1}) to find the orbital basis Green's function as stated in the main text,
\begin{align}
\label{eq:3}
G_{\ell\ell'} (\k,\omega_n) &=\sum_\mu {a_\mu^\ell(\k) a_\mu^{\ell'*}(\k)}G_\mu (\k,\omega_n)\nonumber\\
&= \sum_\mu \frac{a_\mu^\ell(\k) a_\mu^{\ell'*}(\k)}{i \omega_n - \xi_\mu(\k)}.
\end{align}

\section{Quasiparticle description in band space}
At this point, we make a short remark about the implications of quasiparticles in band representation.
Starting from Eq. (\ref{eq:5}), we can transform back to the band basis and obtain the quasiparticle Green's function
\begin{eqnarray}\label{eq:6}
  \tilde G_\nu(\k,\omega_n)&=&\sum_{s,p} { a_\nu^s}^* (k) a_\nu^p (\k) \tilde G_{sp} (\k,\omega_n) \nonumber\\
  &=&\left(\sum_{s,p} |a_\nu^s(\k)|^2 |a_\nu^p(\k)|^2  \sqrt{Z_s}\sqrt{Z_p} \right)   G_\nu (\k,\omega_n)\nonumber\\
  &=&  \tilde Z_\nu(\k) G_\nu(\k,\omega_n) \equiv \tilde G_\nu(\k,\omega_n),
\end{eqnarray}
where $\tilde Z_\nu(\k)\equiv [\sum_s |a_\nu^s(\k)|^2 \sqrt{Z_s}]^2$ are the quasiparticle band weights near the Fermi surface.  If
the point $\k$ on the Fermi surface sheet $\nu$ is dominated by a particular orbital weight $|a_\nu^s(\k)|^2$, the quasiparticle weight for that band will be given predominantly
by $Z_s$.
 Calculating the spectral function from such a Green's function and plotting versus $\k$ at $\omega=0$, one directly sees that part of the
Fermi surface is strongly suppressed in intensity whenever an orbital dominates that has small quasiparticle weight, i.e., is strongly correlated.
In Fig. \ref{fig_pair} we show this effect of the spectral function on the example of our model for FeSe (bulk).

We stress that the approach applied in this paper is phenomenological in the sense that the band renormalizations and the quasiparticle weights are not obtained self-consistently from the same bare interaction parameters. Thus we do not address the problem of how to quantitatively  capture nontrivial self-energy effects and the eventual transition to non-Fermi-liquid behavior with increasing correlations or hole-doping\cite{Yin2011}, but simply rely on a wealth of previous theoretical studies showing the existence of orbital selectivity, and study their influence on the superconducting pairing structure.

\section{Spin-fluctuation pairing: uncorrelated model}
\label{ap_sp_fl}
 Here, we remind the reader of the approach to calculating the gap function in the usual spin fluctuation pairing model\cite{s_graser_09,Kreisel13}.  First,
 local interactions are included via the five-orbital Hubbard-Hund Hamiltionan,
\begin{eqnarray}
	H = H_{0}& + &{U}\sum_{i,\ell}n_{i\ell\uparrow}n_{i\ell\downarrow}+{U}'\sum_{i,\ell'<\ell}n_{i\ell}n_{i\ell'}
	\nonumber\\
	& + & {J}\sum_{i,\ell'<\ell}\sum_{\sigma,\sigma'}c_{i\ell\sigma}^{\dagger}c_{i\ell'\sigma'}^{\dagger}c_{i\ell\sigma'}c_{i\ell'\sigma}\\
	& + & {J}'\sum_{i,\ell'\neq\ell}c_{i\ell\uparrow}^{\dagger}c_{i\ell\downarrow}^{\dagger}c_{i\ell'\downarrow}c_{i\ell'\uparrow} \nonumber, \label{H}
\end{eqnarray}
where the interaction parameters ${U}$, ${U}'$, ${J}$, ${J}'$ are given in the notation of Kuroki \textit{et al.} \cite{Kuroki08} with the choice $U'=U-2J$, $J=J'$, leaving only $U$ and $J/U$ to specify the interactions. Here, $\ell$ is an orbital index with $\ell\in(1,\ldots,5)$ corresponding to the Fe $3d$ orbitals $(d_{xy},d_{x^2-y^2},d_{xz},d_{yz},d_{3z^2-r^2})$.
\label{subsec:spinfluct}
The orbital susceptibility tensor in the normal state is now given as
\begin{align}
	\chi_{\ell_1 \ell_2 \ell_3 \ell_4}^0 (q) & = - \sum_{k,\mu\nu} M_{\ell_1\ell_2\ell_3\ell_4}^{\mu\nu} (\k,\q)
	 G^{\mu}(k+q) G^{\nu} (k),   \label{eqn_supersuscept}
\end{align}
where we have adopted the shorthand $k\equiv (\k,\omega_n)$, and  defined
\begin{equation}
	M_{\ell_1 \ell_2 \ell_3 \ell_4}^{\mu\nu} (\k,\q) = a_\nu^{\ell_4} (\k) a_\nu^{\ell_2,*} (\k) a_\mu^{\ell_1} (\k+\q) a_\mu^{\ell_3,*} (\k+\q).
\end{equation}
The Matsubara sum in Eq.~(\ref{eqn_supersuscept}) is performed  analytically, and we then evaluate $\chi_{\ell_1 \ell_2 \ell_3 \ell_4}^0$ by integrating over the full Brillouin zone.
As noted earlier\cite{Kreisel15}, the Fermi surface nesting condition gives significant contributions to the susceptibility, but finite-energy nesting also contributes.
The spin- ($\chi_1^{\rm RPA}$) and charge-fluctuation ($\chi_0^{\rm RPA}$) parts of the RPA susceptibility for \mbox{$q=(\q,\omega_n=0)$} are now defined within the random phase approximation as
\begin{subequations}
\label{eqn:RPA}
\begin{align}
 \chi_{1\,\ell_1\ell_2\ell_3\ell_4}^{\rm RPA} (\q) &= \left\{ \chi^0 (q) \left[1 -\bar U^s \chi^0 (q) \right]^{-1} \right\}_{\ell_1\ell_2\ell_3\ell_4},\\
 \chi_{0\,\ell_1\ell_2\ell_3\ell_4}^{\rm RPA} (\q) &= \left\{ \chi^0 (q) \left[1 +\bar U^c \chi^0 (q) \right]^{-1} \right\}_{\ell_1\ell_2\ell_3\ell_4}.
\end{align}
\end{subequations}
The total spin susceptibility at $\omega=0$ is then given by the sum
\begin{equation}
	\label{eqn_chisum} \chi (\q) = \frac 12 \sum_{\ell \ell^\prime} \chi_{1\;\ell \ell \ell^\prime\ell^\prime}^{\rm RPA} (\q)\,.
 \end{equation}
 
The interaction matrices $\bar U^s$ and $\bar U^c$ in orbital space are composed of linear combinations of $U,U',J,J'$ and their forms are given, e.g., in Ref.~\onlinecite{a_kemper_10}.
We focus here on the spin-singlet  vertex for pair scattering between bands $\nu$ and $\mu$,  
\begin{eqnarray}
	{\Gamma}_{\nu\mu} (\k,\k') & = & \mathrm{Re}\sum_{\ell_1\ell_2\ell_3\ell_4} a_{\nu}^{\ell_1,*}(\k) a_{\nu}^{\ell_4,*}(-\k) \\
	&&\times {\Gamma}_{\ell_1\ell_2\ell_3\ell_4} (\k,\k') \; a_{\mu}^{\ell_2}(\k') a_{\mu}^{\ell_3}(-\k')\,,\nonumber \label{eq:Gam_ij}
\end{eqnarray}
where $\k$ and $\k'$ are quasiparticle momenta restricted to the pockets $\k \in C_\nu$ and $\k' \in C_\mu$, and is defined in terms of the  
  the orbital space vertex function
\begin{align}
	&{\Gamma}_{\ell_1\ell_2\ell_3\ell_4} (\k,\k') = \left[\frac{3}{2} \bar U^s \chi_1^{\rm RPA} (\k-\k') \bar U^s \right.\,\,\\
	&\,\left. +  \frac{1}{2} \bar U^s - \frac{1}{2}\bar U^c \chi_0^{\rm RPA} (\k-\k') \bar U^c + \frac{1}{2} \bar U^c \right]_{\ell_1\ell_2\ell_3\ell_4}.\nonumber \label{eq:fullGamma}
\end{align}
Using this approximation to the vertex, we now consider the linearized gap equation
 \begin{equation}\label{eqn:gapeqn}
-\frac{1}{V_G}  \sum_\mu\int_{\text{FS}_\mu}dS'\; \Gamma_{\nu\mu}(\k,\k') \frac{ g_i(\k')}{|v_{\text{F}\mu}(\k')|}=\lambda_i g_{i}(\k)
 \end{equation} and solve for the leading eigenvalue $\lambda$ and corresponding eigenfunction $g(\k)$.
Here $v_{\text{F}\mu}(\k')$ is the Fermi velocity of band $\mu$ and the integration is over the Fermi surface $\text{FS}_\mu$.
The eigenfunction $g_i(\k)$ for the leading eigenvalue then determines the symmetry and structure of the leading pairing gap $\Delta(\k)\propto g(\k)$  close to $T_c$.
Finally, the area of the Fermi surface sheets is discretized using a Delaunay triangulation  algorithm that  transforms the integral equation Eq. (\ref{eqn:gapeqn}) into an algebraic matrix equation which is solved numerically. Typically, we use a $\k$-mesh of $80 \times 80 \times 30$ points for the $\k$ integration and totally $\approx 1200$ points on all Fermi sheets for a 3D calculation, while for a 2D calculation the $\k$ mesh is on the order of $100\times 100$ and $\approx 200$ points on all Fermi sheets are required for reasonably converged results.

\begin{figure}[tb]
 \includegraphics[width=0.98\linewidth]{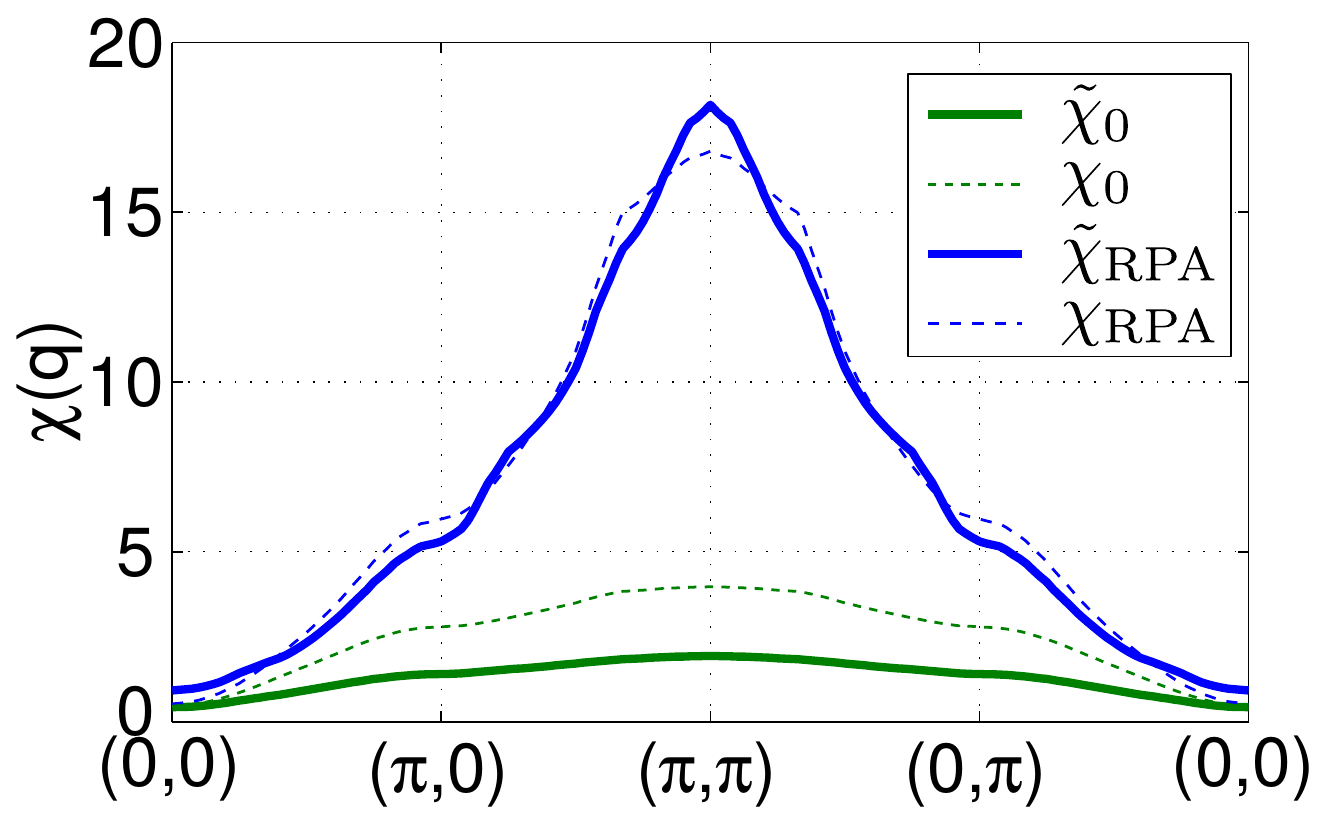}
 \caption{Susceptibility $\tilde \chi$ for our model for the monolayer FeSe as calculated from the orbital selective ansatz using the quasiparticle Green's functions with $\{\sqrt{Z_l}\}=[ 0.4273 ,   0.8000  ,  0.9826 ,   0.9826 , 0.700]$ compared to the conventional calculation ($\chi$), where the interactions have been scaled down.}
 \label{Chi_monolayer_FeSe}
\end{figure}
\begin{figure}[tb]
\includegraphics[width=0.98\linewidth]{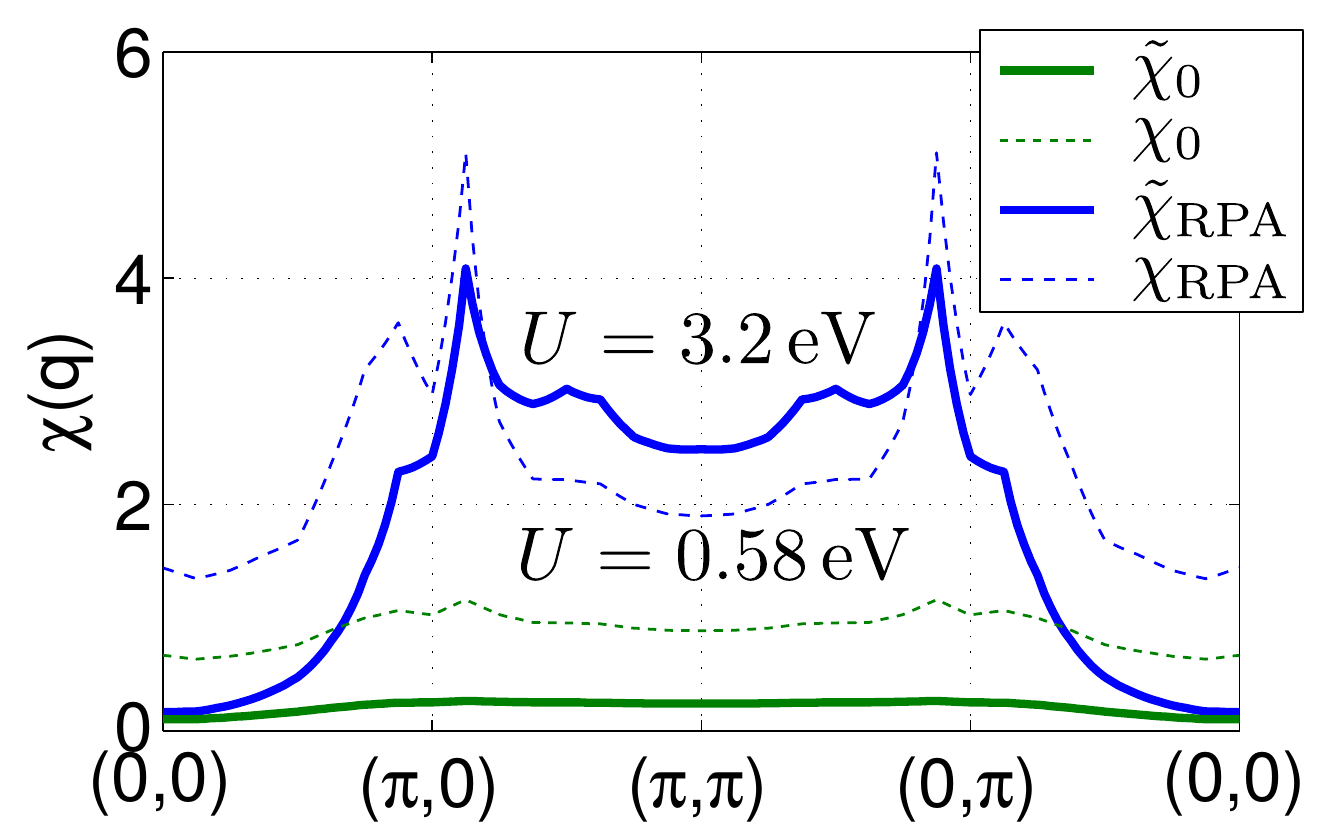}
\caption{Total susceptibility $\chi$  for LiFeAs as calculated from the electronic structure using a 3D model and same quantity $\tilde \chi$ , but calculated using the quasiparticle Green's functions with
$\{\sqrt{Z_l}\}=[0.5493,0.969,0.5952,0.5952,0.9267]$.
}
\label{physical_susc_FeSe_dress}
\end{figure}

\begin{figure*}[tb]
\includegraphics[width=\linewidth]{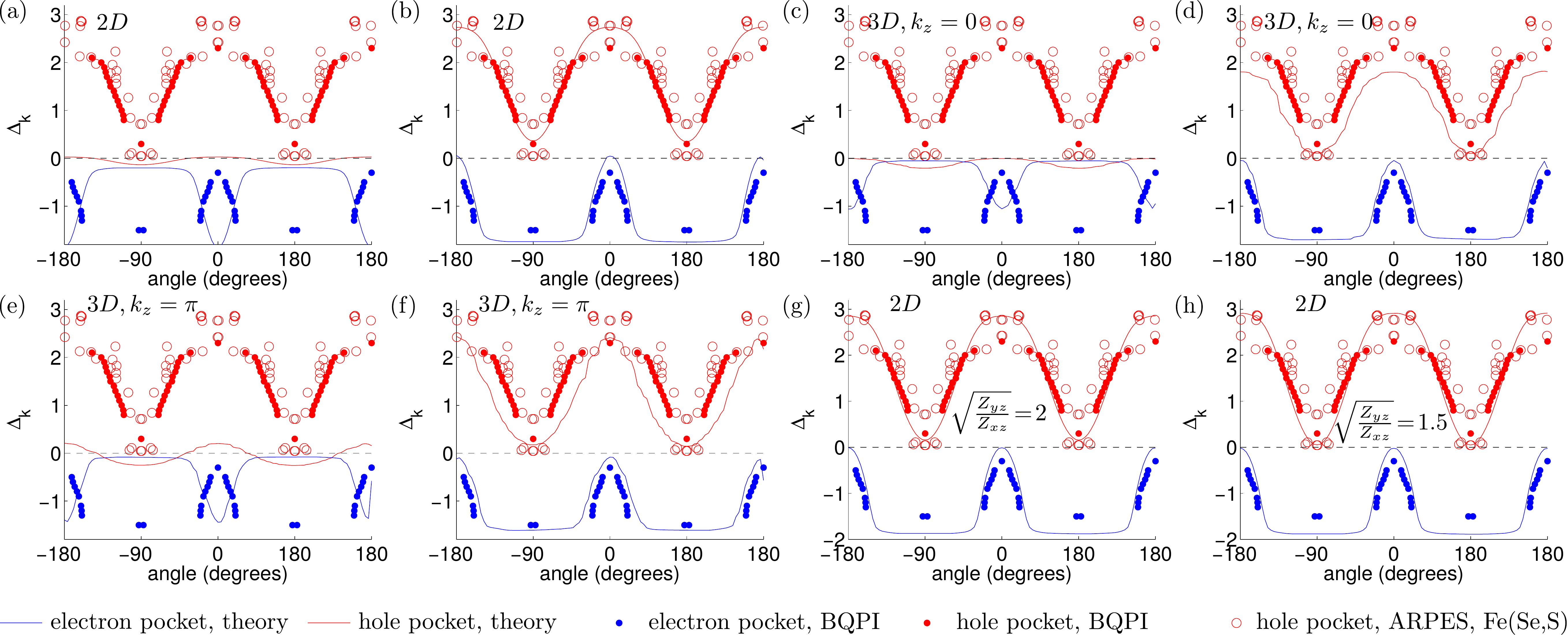}
\caption{Comparison of the calculated gap function for FeSe (bulk) to experimental data from Refs. \cite{Davis_OSP} and \cite{Xu2016}. Calculated gap function from the two-dimensional model at $k_z=0$ with conventional spin-fluctuation pairing and interaction parameters $U=0.33\,\text{eV}$, $J=U/6$, (a), a calculation with the orbitally selective pairing ansatz as described in the main text (b). Since the quasiparticle weights reduce the susceptibility in general, a slightly larger
interaction of $U=0.54\,\text{eV}$ was chosen, while the ratio $J=U/6$ is kept constant. Cuts of the results as shown in the main text for a 3D calculation: (c) $k_z=0$ cut from the conventional spin-fluctuation calculation, (d) the same cut from the orbitally selective ansatz, cuts for $k_z=\pi$ are shown in \mbox{Fig. \ref{Fig_FeSe_bulk}}. Variations of the fits for the 2D model, where the ratio of the quasiparticle weights of the $d_{yz}$ and $d_{xz}$ orbital is constrained to the value as indicated on the figure (g,h). The resulting values are then $\{\sqrt{Z_l}\}=[ 0.2264, 0.9717, 0.4658, 0.9317, 0.6916]$ (g) and $\{\sqrt{Z_l}\}=[ 0.2633, 0.9000, 0.5998, 0.8997, 0.3630]$ (h).}
\label{gap_compare}
\end{figure*}
\section{Spin-fluctuation pairing including quasiparticle weights}
\label{sp_fl_qp}
In this appendix, we show the modified equations for the pairing calculation as outlined above, but including quasiparticle weights from dressed electrons. Taking the ansatz for the dressed Green's function, Eq. (\ref{eq:5}), it is obvious that from Eq. (\ref{eqn_supersuscept}) immediately follows Eq. (\ref{eq_susc}) which is then used in Eqs. (\ref{eqn:RPA}) instead of $\chi_{\ell_1 \ell_2 \ell_3 \ell_4}^0 (q)$ for the dressed quantities. The total susceptibility then reads as
\begin{equation}
\tilde \chi (\q) = \frac 12 \sum_{\ell \ell^\prime} \tilde\chi_{1\;\ell \ell \ell^\prime\ell^\prime}^{\rm RPA} (\q).
\end{equation}
For the FeSe (bulk) model, the total susceptibility is displayed and discussed in the main text, because the quasiparticle weights have a strong effect on the qualitative behavior.
At this point, it is worth mentioning that this is not the case for the model of monolayer FeSe, where the quasiparticle weights are chosen closer to unity (accounting for smaller correlation effects in this material).
In Fig. \ref{Chi_monolayer_FeSe}, it can be seen that the total susceptibility is practically unchanged. Similar conclusions can also be drawn from the comparison of the total susceptibilities for LiFeAs in the uncorrelated and correlated model, see Fig. \ref{physical_susc_FeSe_dress}.
Note that the quasiparticle weights $Z_l$ are consistent with DMFT results where it is found that $t_{2g}$ orbitals are strongly correlated with $d_{xy}$ strongest, and components of the susceptibility get suppressed ($d_{xy}$ strongest)\cite{Nourafkan16}.

The equation
\begin{align}
	&{\tilde\Gamma}_{\ell_1\ell_2\ell_3\ell_4} (\k,\k') = \left[\frac{3}{2} \bar U^s \tilde\chi_1^{\rm RPA} (\k-\k') \bar U^s \right.\,\,\\
	&\,\left. +  \frac{1}{2} \bar U^s - \frac{1}{2}\bar U^c \tilde\chi_0^{\rm RPA} (\k-\k') \bar U^c + \frac{1}{2} \bar U^c \right]_{\ell_1\ell_2\ell_3\ell_4}\nonumber \label{eq:fullGammadress}
\end{align}
for the orbital space vertex function is basically unchanged except for the addition of the tilde.
In the construction of the pair scattering vertex, additional quasiparticle weights enter from the replacement $c^\dagger_\ell(\k) \rightarrow\sqrt{Z_\ell} c^\dagger_\ell (\k)$ such that it reads
\begin{align}
	&{\tilde\Gamma}_{\nu\mu} (\k,\k')  = \mathrm{Re}\sum_{\ell_1\ell_2\ell_3\ell_4} \sqrt{Z_{\ell_1}}\sqrt{Z_{\ell_4}} a_{\nu}^{\ell_1,*}(\k) a_{\nu}^{\ell_4,*}(-\k) \nonumber\\
	&\hspace{0.6cm}\times {\tilde\Gamma}_{\ell_1\ell_2\ell_3\ell_4} (\k,\k') \; \sqrt{Z_{\ell_2}}\sqrt{Z_{\ell_3}} a_{\mu}^{\ell_2}(\k') a_{\mu}^{\ell_3}(-\k')\, 
\end{align}
and enters Eq. (\ref{eqn:gapeqn}) instead of $\Gamma_{\nu\mu}(\k,\k')$.

\section{Comparison of 2D calculations and 3D calculations}

In the present paper, we discuss three different physical systems, two of them parametrized using a band structure including a $k_z$ dispersion as well.
As noted already earlier, the susceptibility as calculated from a 3D model (with weak dispersion in $k_z$ direction) shows only very small dependence on $k_z$\cite{Wang13}.
Conclusions similar to the ones in the main text can also be drawn in a two-dimensional calculation, where the initial band structure is just the one at $k_z=0$. Taking the same interaction parameters and quasiparticle weights, one obtains qualitative similar results as for the 3D calculation. This is expected since the electronic structure is found to be quasi-two-dimensional, and especially since the susceptibility and thus the pairing interaction have little dependence on $q_z$. Differences in the relative magnitudes of the gap functions on the individual pockets can, however, arise due to the variation of the Fermi velocities as a function of $k_z$, e.g. the weight at $k_z=0$ as included in a 2D calculation is not just the average of the partial contributions to the density of states from different $k_z$\cite{Wang13}. In the solution of the linearized gap equation, this can increase the gap on individual pockets\cite{Wang13} or reduce the gap as seen on the $\alpha$ pocket for the 3D calculation in Fig. \ref{gap_compare} (d).
Overall, the variation of the results is small and mostly of quantitative nature rather than qualitative.
We note that the Fermi surface properties can still strongly influence the actual superconducting order parameter in such a calculation even if the pairing interaction itself has negligible variation in $q_z$.
This will occur in a 2D calculation for the LiFeAs model where the Fermi surface is different at cuts in $k_z=0$ and $k_z=\pi$ because of the closed $\alpha$ pocket. Because of this, we have not  considered any results of a 2D calculation for this model further.
Finally, we present results for the gap structure obtained from a fit where the relative magnitudes of the quasiparticle weights of the $d_{xz}$ and $d_{yz}$ orbital are kept fixed. Even when lowering the ratio between those, the agreement is still good [see Fig. \ref{gap_compare} (g-h)], but not allowing a larger quasiparticle weight in the $d_{yz}$ orbital does not yield an agreement (not shown).

\end{document}